\documentclass[runningheads]{llncs}
\usepackage{graphicx}
\usepackage{import}
\usepackage{amsmath,graphicx}
\usepackage{times}
\usepackage[dvipsnames,table,x11names]{xcolor}

\usepackage{tabularx,booktabs}
\usepackage{arydshln}
\usepackage{tabu}
\usepackage{adjustbox}
\usepackage{enumerate,multicol}
\usepackage{multirow}
\usepackage{multicol}
\usepackage{enumitem}
\usepackage{amsmath,array,graphicx}
\usepackage{gensymb}
\usepackage{float}
\usepackage{caption}
\usepackage{subcaption}
\usepackage{balance}
\usepackage{xspace}
\usepackage{amssymb}
\usepackage{pifont}
\usepackage{subcaption}
\usepackage{tabularx}
\usepackage{hyperref}
\usepackage{amsmath}

\hypersetup{
    colorlinks,
    linkcolor={red!50!black},
    citecolor={blue!50!black},
    urlcolor={black}
}

\usepackage{array}
\newcolumntype{P}[1]{>{\centering\arraybackslash}p{#1}}

\newcommand{\xmark}{\ding{55}}

\newcommand{\beginsupplement}{%
        \setcounter{table}{0}
        \renewcommand{\thetable}{S\arabic{table}}%
        \setcounter{figure}{0}
        \renewcommand{\thefigure}{S\arabic{figure}}%
     }

\usepackage{times}

\usepackage{soul}
\usepackage{url}
\usepackage[]{hyperref}
\usepackage[utf8]{inputenc}
\usepackage{amsmath}
\usepackage{booktabs}
\usepackage{pgfplots}
\usepackage{pgfplotstable}
\usetikzlibrary{pgfplots.groupplots}
\usepgfplotslibrary{colorbrewer}
\pgfplotsset{/pgfplots/error bars/error bar style={black,thick}}

\pgfplotsset{compat=1.11,
        /pgfplots/ybar legend/.style={
        /pgfplots/legend image code/.code={%
        \draw[##1,/tikz/.cd,bar width=3pt,yshift=-0.2em,bar shift=0pt]
                plot coordinates {(0cm,0.8em)};},
},}
\begin{document}
\title{LensID: A CNN-RNN-Based Framework \\Towards Lens Irregularity Detection \\in Cataract Surgery Videos\thanks{This work was funded by the FWF Austrian Science Fund under grant P 31486-N31.}}
\titlerunning{LensID: A CNN-RNN-Based Framework Towards Lens Irregularity Detection}
%

\author{Negin Ghamsarian\inst{1} \and
Mario Taschwer\inst{1} \and
Doris Putzgruber-Adamitsch\inst{2} \and
Stephanie Sarny\inst{2} \and
Yosuf El-Shabrawi\inst{2} \and
Klaus Schoeffmann \inst{1}}

\authorrunning{N. Ghamsarian et al.}
%

\institute{Department of Information Technology, Alpen-Adria-Universit\"at Klagenfurt \and
Department of Ophthalmology, Klinikum Klagenfurt\\
\email{negin@itec.aau.at}}
\maketitle              
\begin{abstract}
A critical complication after cataract surgery is the dislocation of the lens implant leading to vision deterioration and eye trauma. In order to reduce the risk of this complication, it is vital to discover the risk factors during the surgery. However, studying the relationship between lens dislocation and its suspicious risk factors using numerous videos is a time-extensive procedure. Hence, the surgeons demand an automatic approach to enable a larger-scale and, accordingly, more reliable study. In this paper, we propose a novel framework as the major step towards lens irregularity detection. In particular, we propose (I) an end-to-end recurrent neural network to recognize the lens-implantation phase and (II) a novel semantic segmentation network to segment the lens and pupil after the implantation phase. The phase recognition results reveal the effectiveness of the proposed surgical phase recognition approach. Moreover, the segmentation results confirm the proposed segmentation network's effectiveness compared to state-of-the-art rival approaches.

\end{abstract}
\section{Introduction}
\label{sec: Introduction}

Cataract refers to the eye's natural lens, having become cloudy and causing vision deterioration. Cataract surgery is the procedure of restoring clear eye vision via cataract removal followed by artificial lens implantation. This operation is conducted with the aid of a binocular microscope providing a 3D magnified and illuminated image of the eye. The camera mounted on the microscope records and stores the whole surgery for postoperative objectives.

Over several years, there have been numerous advances in surgical techniques, tools, and instruments in ophthalmic surgeries. Such advances resulted in decreasing the risk of severe intraoperative and postoperative complications. Still, there are many ongoing research efforts to prevent the current implications during and after surgery. A critical issue in cataract surgery that has not yet been addressed is intraocular lens (IOL) dislocation. This complication leads to various human sight issues such as vision blur, double vision, or vision inference as observing the lens implant edges. Intraocular inflammation, corneal edema, and retinal detachment are some other consequences of lens relocation. Since patient monitoring after the surgery or discharge is not always possible, the surgeons seek ways to diagnose evidence of potential irregularities that can be investigated during the surgery. 

Recent studies show that particular intraocular lens characteristics can contribute to lens dislocation after the surgery~\cite{IIOL}. Moreover, the expert surgeons argue that there can be a direct relationship between the overall time of lens unfolding and the risk of lens relocation after the surgery. Some surgeons also hypothesize that severe lens instability during the surgery is a symptom of lens relocation. To discover the potential correlations between lens relocation and its possibly contributing factors, surgeons require a tool for systematic feature extraction. Indeed, an automatic approach is required for (i) detecting the lens implantation phase to determine the starting time for lens statistics' computation and (ii) segmenting the lens and pupil to compute the lens statistics over time. The irregularity-related statistics can afterward be extracted by tracking the lens's relative size (normalized by the pupil's size) and relative movements (by calibrating the pupil). 
Due to the dearth of computing power in the operation rooms, automatic phase detection and lens/pupil segmentation on the fly is not currently achievable. Alternatively, this analysis can be performed in a post hoc manner using recorded cataract surgery videos. The contributions of this paper are: 
\begin{enumerate}
    \item We propose a novel CNN-RNN-based framework for evaluating lens unfolding delay and lens instability in cataract surgery videos. 
    \item We propose and evaluate a recurrent convolutional neural network architecture to detect the ``implantation phase'' in cataract surgery videos.
    \item We further propose a novel semantic segmentation network architecture termed as \textit{AdaptNet}\footnote[1]{The PyTorch implementation of AdaptNet is publicly available at https://github.com/Negin-Ghamsarian/AdaptNet-MICCAI2021.}, that can considerably improve the segmentation performance for the intraocular lens (and pupil) compared to ten rival state-of-the-art approaches.
    \item We introduce three datasets for phase recognition, pupil segmentation, and lens segmentation that are publicly released with this paper to support reproducibility and allow further investigations for lens irregularity detection\footnote[2]{http://ftp.itec.aau.at/datasets/ovid/LensID/}.
\end{enumerate}
\section{Related Work}
\label{sec: relatedwork}
Since this work involves \textit{phase recognition} and \textit{semantic segmentation}, we briefly review the state-of-the-art approaches related to the mentioned subjects.

Phase recognition in surgical videos has experienced three stages of development. In the first stage, hand-engineered features such as color, shape, and instrument presence information~\cite{FRHS,RTS,RFPD} are exploited as the input to classical machine learning approaches. Using Conditional Random Fields (CRF), Random Forests, and Hidden Markov Models (HMM), the corresponding phase of each feature vector is classified ~\cite{SGC,RTS,RFPD}. Because of pruning to suboptimal results, the classical approaches are replaced with convolutional neural networks in the second stage~\cite{EndoNet}. To improve the classification accuracy by taking advantage of the temporal information, the third generation of phase recognition approaches adopt LSTM~\cite{LSTM} or GRU~\cite{GRU} or bidirectional recurrent layers~\cite{DPS,SV-RCNet,RDCSV}.

Since semantic segmentation plays a prominent role in medical image analysis, considerable attention has been devoted to this subject in recent years. Specifically, U-Net~\cite{U-Net}, which takes advantage of skip connections between symmetric layers in encoder and decoder, demonstrated pioneering results in medical image segmentation. Many approaches based on U-Net are proposed to improve its segmentation accuracy or address its weaknesses~\cite{CE-Net,MultiResUNet,dU-Net,CPFNet,UNet++}. UNet++~\cite{UNet++} ensembles  varying-depth U-Nets to deal with network depth optimization. CPFNet~\cite{CPFNet} adopts a scale-aware module by fusing the output feature maps of atrous convolutions~\cite{Atrous} with different dilation rates. MultiResUNet~\cite{MultiResUNet} fuses the feature maps coming from sequential convolutions as an effective alternative to convolutional layers with large filters and atrous convolutions. Many other attention modules are proposed to boost segmentation accuracy for different medically-relevant content such as surgical instruments~\cite{RAUNet,BARNet,PAANet}, liver lesion~\cite{FED-Net}, and general medical image segmentation~\cite{CE-Net}.
\section{Methodology}
\label{sec: Methodology}
\begin{figure}[!tb]
    \centering
    \includegraphics[width=1\textwidth]{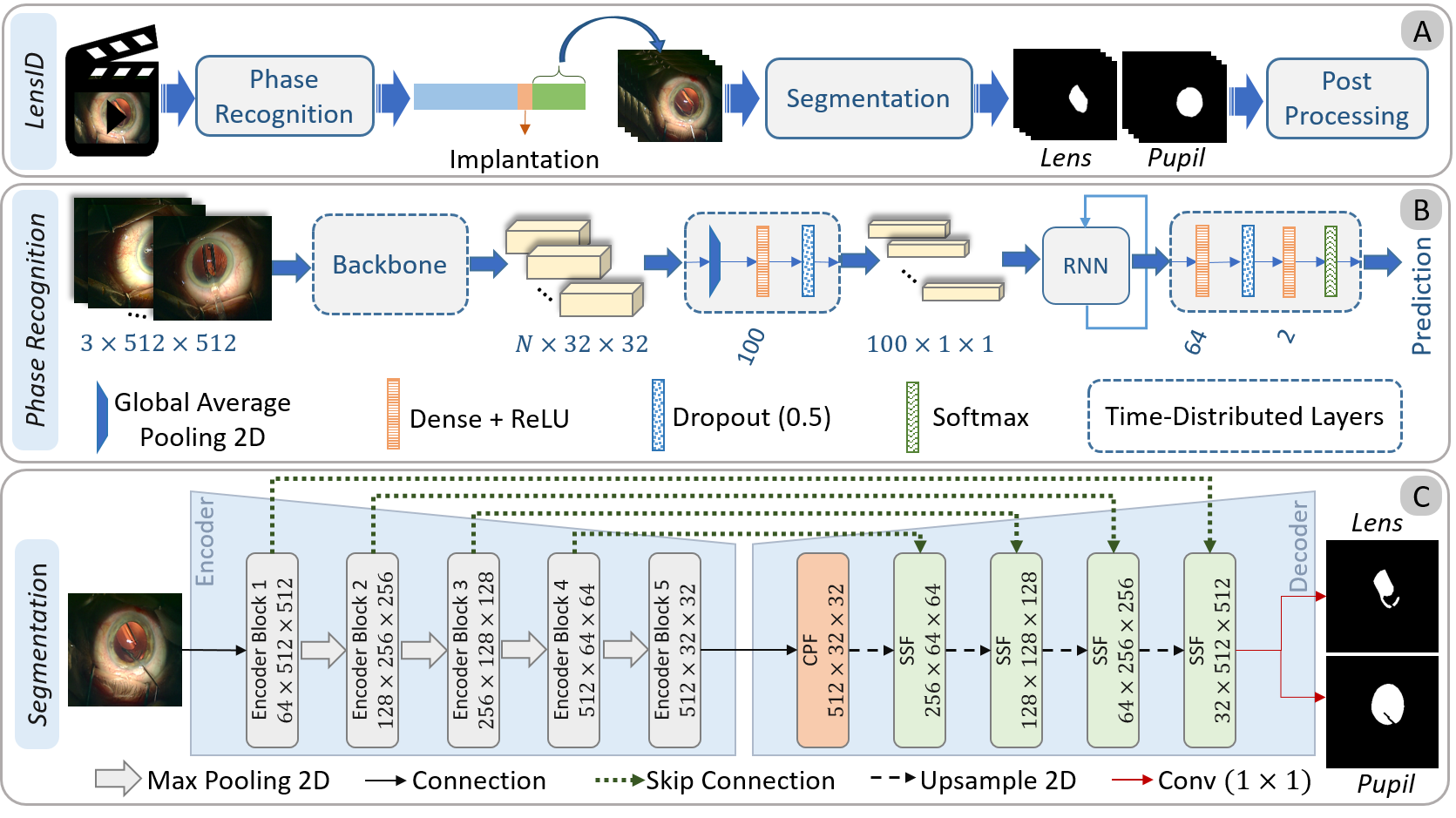}
    \caption{The block diagram of \textit{LensID} and the architecture of \textit{Phase Recognition} and \textit{Semantic Segmentation} networks.}
    \label{fig: BD1}
\end{figure}
Fig.~\ref{fig: BD1} demonstrates the block diagram of \textit{LensID} and the network architecture of the phase recognition and segmentation steps. As the first step towards lens irregularity detection, we adopt a recurrent convolutional network (Fig.~\ref{fig: BD1}-B) to detect the lens implantation phase (the temporal segment in which the lens implantation instrument is visible). We start segmenting the artificial lens and pupil exactly after the lens implantation phase using the proposed semantic segmentation network (Fig.~\ref{fig: BD1}-C). The pupil and lens segmentation results undergo some post-processing approaches to compute lens instability and lens unfolding delay. More precisely, we draw the smallest convex polygon surrounding pupil's and lens' masks using binary morphological operations. For lens instability, we use the normalized distance between the lens and pupil centers. For lens unfolding, we track the lens' area over time, considering its relative position.
\paragraph{\textbf{Phase Recognition. }}
As shown in Fig.~\ref{fig: BD1}-B, we use a pre-trained backbone followed by global average pooling to obtain a feature vector per each input frame. These features undergo a sequence of Dense, Dropout, and ReLU layers to extract higher-order semantic features. A recurrent layer with five units is then employed to improve the feature representation by taking advantage of temporal dependencies. These features are then fed into a sequence of layers to finally output the predicted class for each input frame.

\begin{figure}[!tb]
    \centering
    \includegraphics[width=1 \textwidth]{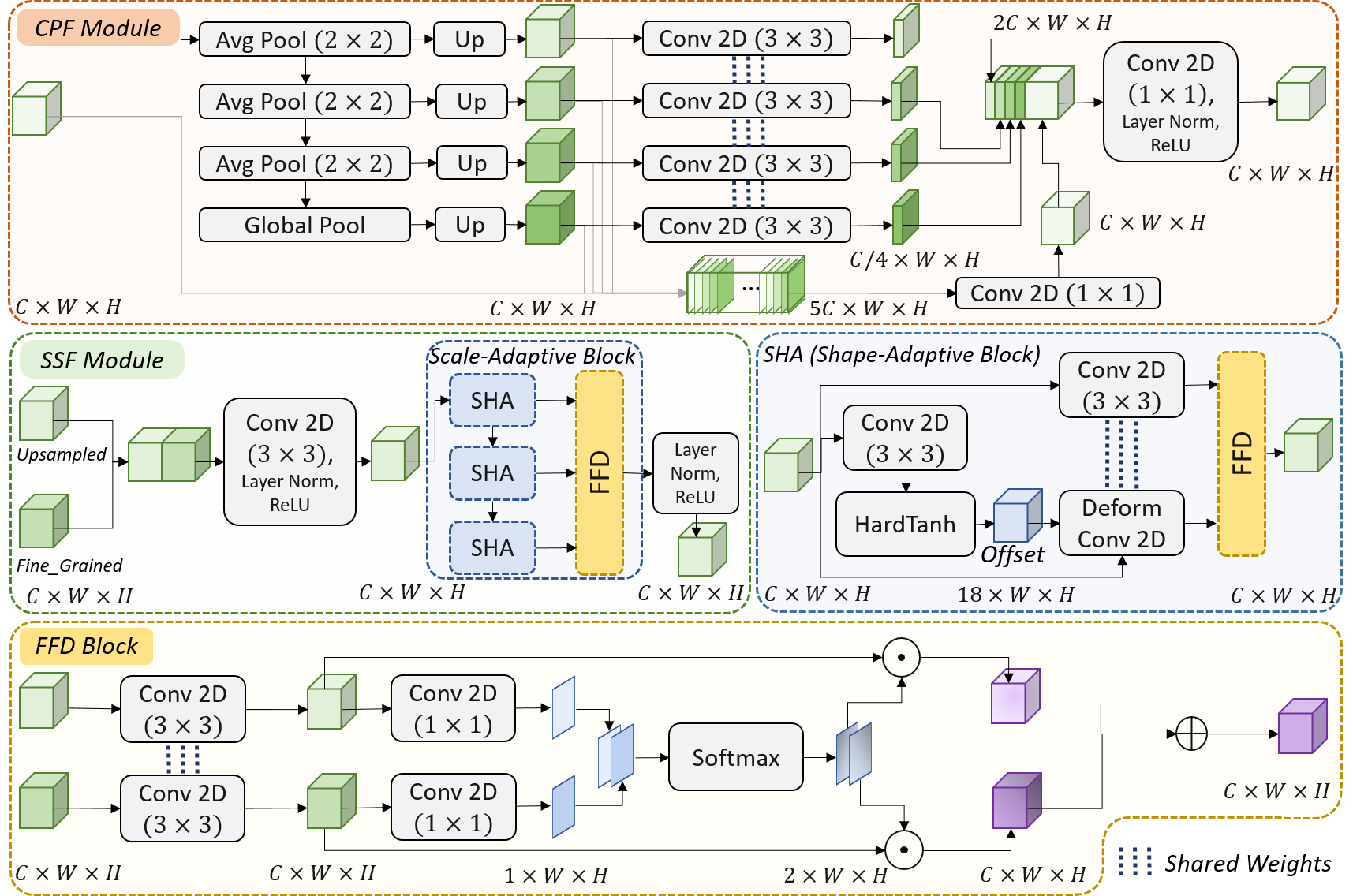}
    \caption{The detailed architecture of the \textit{CPF} and \textit{SFF} modules of AdaptNet.}
    \label{fig: AdaptNet}
\end{figure}

\paragraph{\textbf{Lens \& Pupil Segmentation. }}
In cataract surgery, a folded artificial lens is implanted inside the eye. The lens is transparent and inherits the pupil's color after implantation. Moreover, it is usually being unfolded very fast (sometimes with the help of an instrument). The transparency and unpredictable formation of this object, as well as occlusion, defocus blur, and motion blur~\cite{DCS}, make lens segmentation and tracking more challenging. Hence, we require a semantic segmentation network that can be adapted to the changes in the artificial lens's shape and scale.
We adopt a U-Net-based encoder-decoder architecture for the proposed semantic segmentation network termed as AdaptNet. AdaptNet consists of three main components: encoder, \textit{cascade pooling fusion (CPF)} module, and \textit{shape/scale-adaptive feature fusion (SSF)} module. We use the VGG16 network as the encoder network.
The encoder's output feature map is fed into the CPF module to enhance the feature representation using pyramid features. This feature map is then fed into a sequence of SSF modules, which decode low-resolution semantic features.

As shown in Fig.~\ref{fig: AdaptNet}, the CPF module applies a sequence of three average pooling layers (with a stride of two pixels) followed by a global average pooling layer to the input features. The obtained feature maps are upsampled to the original size of the input and concatenated together with the input feature map in a depth-wise manner. Each group of five channels in the generated feature map undergoes a distinct convolution for intra-channel feature refinement (which is performed using a convolutional layer with $C$ groups). Besides, the upsampled features are mapped into a smaller channel space while extracting higher-order semantic features using convolutional layers with shared weights. The obtained features are concatenated with the intra-channel refined features and undergo a convolutional layer for inter-channel feature refinement.

The \textit{SSF} module starts with concatenating the upsampled semantic feature map with the fine-grained feature map coming from the encoder. The concatenated feature map is fed into a sequence of convolutional,  layer normalization, and ReLU layers for feature enhancement and dimensionality reduction. The resulting features are fed into the \textit{scale-adaptive block}, which aims to fuse the features coming from cascade convolutional blocks. This succession of convolutional layers with small filter sizes can factorize the large and computationally expensive receptive fields~\cite{MultiResUNet}. Moreover, the fusion of these successive feature maps can play the role of scale-awareness for the network. The \textit{shape-adaptive (SHA) block} is responsible for fusing the resulting feature maps of deformable and structured convolutions. At first, a convolutional layer followed by a hard tangent hyperbolic function is employed to produce the offsets for the deformable convolutional layer~\cite{DeformConv}. The input features are also fed into a regular convolutional layer that shares the weights with the deformable layer for structured-feature extraction. These features are then fused to induce the awareness of shape and deformation to the network. 

The \textit{feature fusion decision (FFD)} block inspired by CPFNet~\cite{CPFNet} accounts for determining the importance of each input feature map in improving semantic features. Fig.~\ref{fig: AdaptNet} shows the \textit{FFD Block} in the case of two input branches. At first, shared convolutional layers are applied to the input feature maps to extract the shared semantic features. The resulting feature maps undergo shared pixel-wise convolutions to produce the pixel-wise attention maps. The concatenated attention maps are fed into a softmax activation layer for normalization. The obtained features are used as pixel-wise weights of the shared-semantic feature maps. The shape/scale adaptive features are computed as the sum of pixel-wise multiplications ($\odot$) between the normalized attention maps and their corresponding semantic feature maps.
\section{Experimental Setup}
\label{sec: Experimental Settings}
We use three datasets for this study: (i)~a large dataset containing the annotations for the lens implantation phase versus the rest of phases from 100 videos/operations of cataract surgery, (ii)~a dataset containing the lens segmentation of 401 frames from 27 videos (292 images from 21 videos for training, and 109 images from six videos for testing), and (iii)~a dataset containing the pupil segmentation of 189 frames from 16 videos (141 frames from 13 videos for training, and 48 frames from three videos for testing). Regarding the phase recognition dataset, since lens implantation is a very short phase (around four seconds) compared to the whole surgery (seven minutes on average), creating a balanced dataset that can cover the entire content of videos from the ``Rest" class is quite challenging. Hence, we propose a video clip generator that can provide diverse training sequences for the recurrent neural network by employing stochastic functions. At first, 12 three-second video clips with overlapping frames are extracted from the implantation phase of each cataract surgery video. Besides, the video segments before and after the implantation phase are divided into eight and four video clips, respectively (these clips have different lengths depending on the length of the input video). Accordingly, we have a balanced dataset containing 2040 video clips from 85 videos for training and 360 video clips from the other 15 videos for testing. For each training example, the video generator uses a stochastic variable to randomly select a three-second clip from the input clip. We divide this clip into $N$ sub-clips, and $N$ stochastic variables are used to randomly select one frame per sub-clip (in our experiments, $N$ is set to five to reduce computational complexity and avoid network overfitting). 

We compare the segmentation accuracy of the proposed approach (AdaptNet) with ten state-of-the-art approaches including UNet++ (and UNet++\slash DS)~\cite{UNet++}, MultiResUNet~\cite{MultiResUNet}, CPFNet~\cite{CPFNet}, dU-Net~\cite{dU-Net}, CE-Net~\cite{CE-Net}, FEDNet~\cite{FED-Net}, PSPNet~\cite{PSPNet}, SegNet~\cite{SegNet}, and U-Net~\cite{U-Net}. It should be mentioned that the rival approaches employ different backbone networks, loss functions (cross entropy or cross entropy log Dice), and upsampling methods (bilinear, transposed convolution, pixel-shuffling, or max unpooling).

For phase recognition, all networks are trained for 20 epochs. The initial learning rate for these networks is set to 0.0002 and 0.0004 for the networks with VGG19 and Resnet50 backbones, respectively, and halved after ten epochs. Since the segmentation networks used for evaluations have different depths, backbones, and the number of trainable parameters, all networks are trained with three different initial learning rates ($lr_0\in\{0.0005, 0.001, 0.002\}$). For each network, the results with the highest Dice coefficient are listed. All segmentation networks are trained for 30 epochs, and the learning rate is decreased by a factor of 0.8 in every other epoch.  To prevent overfitting and improve generalization performance, we have used motion blur, Gaussian blur, random contrast, random brightness, shift, scale, and rotation for data augmentation. The backbones of all networks evaluated for phase recognition and lens/pupil semantic segmentation are initialized with ImageNet~\cite{imageNet} weights. The size of input images to all networks is set to $512\times 512\times 3$.
The loss function for the phase recognition network is set to \textit{Binary Cross Entropy}. For the semantic segmentation task, we adopt a loss function consisting of categorical cross entropy and logarithm of soft Dice coefficient as follows (in Eq.~\eqref{eq:loss}, \textit{CE} stands for \textit{Cross Entropy}, and $\mathcal{X}_{Pred}$ and $\mathcal{X}_{True}$ denote the predicted and ground-truth segmentation images, respectively. Besides, we use a Dice smoothing factor equal to 1, and set $\lambda=0.8$ in our experiments):
\begin{align}
\mathcal{L} = \lambda\times CE(\mathcal{X}_{Pred},\mathcal{X}_{True}) - (1-\lambda)\times \log_2 Dice(\mathcal{X}_{Pred},\mathcal{X}_{True})
\label{eq:loss}
\end{align}
To evaluate the performance of phase recognition networks, we use \textit{Precision}, \textit{Recall}, \textit{F1-Score}, and \textit{Accuracy}, which are the common classification metrics. The semantic segmentation performance is evaluated using \textit{Dice coefficient} and \textit{Intersection over Union (IoU)}.
\section{Experimental Results and Discussion}
\label{sec: Experimental Results}
\begin{table}[t!]
\renewcommand{\arraystretch}{1}
\caption{Phase recognition results of the end-to-end recurrent convolutional networks.}
\label{tab:RNN}
\centering

\begin{tabular}{lP{1.25cm}P{1.25cm}P{1.25cm}P{1.25cm}P{1.25cm}P{1.25cm}P{1.25cm}P{1.25cm}}
\specialrule{.12em}{.05em}{.05em}
&\multicolumn{4}{c}{Backbone: VGG19}&\multicolumn{4}{c}{Backbone: ResNet50}\\\cmidrule(lr){2-5}\cmidrule(lr){6-9}
RNN&Precision&Recall&F1-Score&Accuracy&Precision&Recall&F1-Score&Accuracy\\\specialrule{.12em}{.05em}{.05em}
GRU&0.97&0.96&0.96&0.96&0.9&0.94&0.94&0.94\\
LSTM&0.98&0.98&0.98&0.98&0.96&0.96&0.96&0.96\\
BiGRU&0.97&0.96&0.96&0.96&0.95&0.95&0.95&0.95\\
\rowcolor{lightgray} BiLSTM&1.00&1.00&1.00&1.00&0.98&0.98&0.98&0.98\\ \specialrule{.12em}{.05em}{.05em}
\end{tabular}

\end{table}

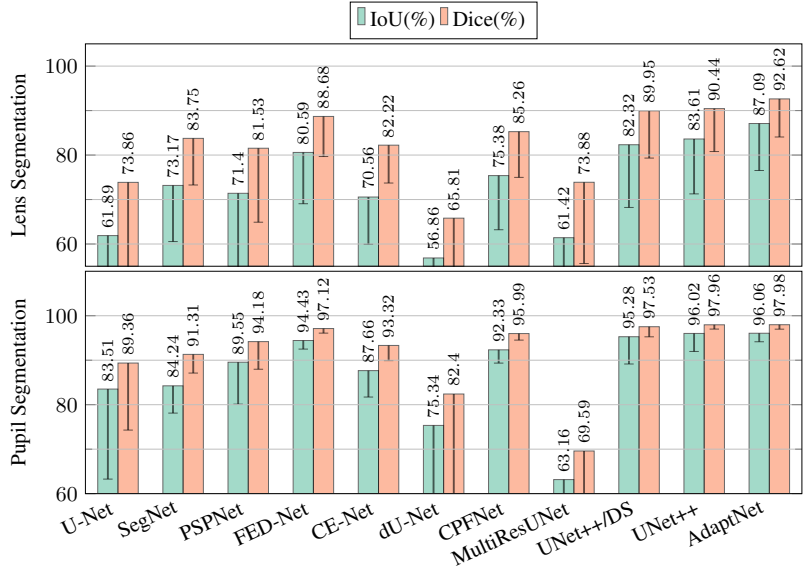
\begin{figure*}[t!]
\begin{tabular}{c}
\begin{subfigure}{1\textwidth}\vspace{-0.5\baselineskip}
  \centering
  \begin{adjustbox}{width=0.87\textwidth}

\begin{tikzpicture}
\begin{axis}[
ybar=0pt,
axis on top,
bar width=0.38cm,
width=\textwidth,
height=.4\textwidth,
ylabel=Lens Segmentation,
enlarge x limits=0.05,
every axis x label/.append style={xshift=0.2cm},
symbolic x coords={U-Net, SegNet, PSPNet, FED-Net, CE-Net, dU-Net, CPFNet, MultiResUNet, UNet++\slash DS, UNet++, AdaptNet},
xtick=\empty,
nodes near coords align={vertical},
cycle list/Set2,
ymin=55,ymax=105,
ymajorgrids = true,
yminorgrids = true,
bar width=0.3cm,
nodes near coords,
minor ytick={70,90},
label style={font=\small},
tick label style={font=\small},
major x tick style = transparent,
x tick label style={rotate=45,anchor=east},
legend style={at={(0.5,+1.19)},
	            anchor=north,legend columns=6},
every node near coord/.append style={rotate=90, anchor=west, font=\scriptsize, color = black, opacity=1}	            
]
\addplot+[style={draw=black,solid,fill,opacity=0.6,
}, 
             error bars/.cd, 
             y dir=minus,y explicit]
             coordinates {
                  (U-Net,61.89) +- (0,20.93)
                  (SegNet,73.17) +- (0,12.62)
                  (PSPNet,71.40) +- (0,18.92)
                  (FED-Net,80.59) +- (0,11.53)
                  (CE-Net,70.56) +- (0,10.62)
                  (dU-Net,56.86) +- (0,32.22)
                  (CPFNet,75.38) +- (0,12.17)
                  (MultiResUNet,61.42) +- (0,19.91)
                  (UNet++\slash DS,82.32) +- (0,14.10)
                  (UNet++,83.61) +- (0,12.34)
                  (AdaptNet,87.09) +- (0,10.56)}; 
                  
\addplot+[style={draw=black,solid,fill,opacity=0.6,
}, 
             error bars/.cd, 
             y dir=minus,y explicit]
             coordinates {
                  (U-Net,73.86) +- (0,20.39)
                  (SegNet,83.75) +- (0,10.48)
                  (PSPNet,81.53) +- (0,16.63)
                  (FED-Net,88.68) +- (0,9.01)
                  (CE-Net,82.22) +- (0,8.51)
                  (dU-Net,65.81) +- (0,32.19)
                  (CPFNet,85.26) +- (0,10.28)
                  (MultiResUNet,73.88) +- (0,18.26)
                  (UNet++\slash DS,89.95) +- (0,10.63)
                  (UNet++,90.44) +- (0,9.65)
                  (AdaptNet,92.62) +- (0,8.56)}; 
\legend{IoU(\%),Dice(\%)}
\end{axis}
\end{tikzpicture}

\end{adjustbox}
\end{subfigure} \\
\begin{subfigure}{1\textwidth}\vspace{-0.2\baselineskip}
  \centering
  \begin{adjustbox}{width=0.87\textwidth}
  \begin{tikzpicture}
\begin{axis}[
ybar=0pt,
axis on top,
bar width=0.38cm,
width=\textwidth,
height=.4\textwidth,
ylabel=Pupil Segmentation,
enlarge x limits=0.05,
every axis x label/.append style={xshift=0.2cm},
symbolic x coords={U-Net, SegNet, PSPNet, FED-Net, CE-Net, dU-Net, CPFNet, MultiResUNet, UNet++\slash DS, UNet++, AdaptNet},
nodes near coords align={vertical},
cycle list/Set2,
ymin=60,ymax=110,
ymajorgrids = true,
yminorgrids = true,
bar width=0.3cm,
nodes near coords,
minor ytick={70,90},
ticklabel style = {font=\small},
major x tick style = transparent,
x tick label style={rotate=25,anchor=east},
legend style={at={(0.5,+1.22)},
	            anchor=north,legend columns=6},
every node near coord/.append style={rotate=90, anchor=west, font=\scriptsize, color = black, opacity=1}	            
]
\addplot+[style={draw=black,solid,fill,opacity=0.6,
}, 
             error bars/.cd, 
             y dir=minus,y explicit]
             coordinates {
                  (U-Net,83.51) +- (0,20.24)
                  (SegNet,84.24) +- (0,6.13)
                  (PSPNet,89.55) +- (0,9.43)
                  (FED-Net,94.43) +- (0,1.93)
                  (CE-Net,87.66) +- (0,5.94)
                  (dU-Net,75.34) +- (0,27.60)
                  (CPFNet,92.33) +- (0,2.98)
                  (MultiResUNet,63.16) +- (0,36.90)
                  (UNet++\slash DS,95.28) +- (0,6.13)
                  (UNet++,96.02) +- (0,4.05)
                  (AdaptNet,96.06) +- (0,1.91)}; 
                  
\addplot+[style={draw=black,solid,fill,opacity=0.6,
}, 
             error bars/.cd, 
             y dir=minus,y explicit]
             coordinates {
                  (U-Net,89.36) +- (0,15.07)
                  (SegNet,91.31) +- (0,4.20)
                  (PSPNet,94.18) +- (0,6.21)
                  (FED-Net,97.12) +- (0,1.03)
                  (CE-Net,93.32) +- (0,3.41)
                  (dU-Net,82.40) +- (0,22.62)
                  (CPFNet,95.99) +- (0,1.47)
                  (MultiResUNet,69.59) +- (0,34.93)
                  (UNet++\slash DS,97.53) +- (0,2.27)
                  (UNet++,97.96) +- (0,0.96)
                  (AdaptNet,97.98) +- (0,1.00)}; 
\end{axis}
\end{tikzpicture}

\end{adjustbox}
\end{subfigure}
\end{tabular}
\caption{Quantitative comparison of segmentation results for the proposed approach (AdaptNet) and rival approaches.}
\label{fig:Dice-IoU}
\end{figure*}

Table~\ref{tab:RNN} compares the classification reports of the proposed architecture for phase recognition considering two different backbone networks and four different recurrent layers. Thanks to the large training set and taking advantage of recurrent layers, all networks have shown superior performance in classifying the implantation phase versus other phases. However, the LSTM and bidirectional LSTM (BiLSTM) layers have shown better performance compared to GRU and BiGRU layers, respectively. Surprisingly, the network with a VGG19 backbone and BiLSTM layer has achieved 100\% accuracy in classifying the test clips extracted from the videos which are not used during training.
Fig.~\ref{fig:Dice-IoU} compares the segmentation results (mean and standard deviation of IoU and Dice coefficient) of AdaptNet and ten rival state-of-the-art approaches. Overall, it can be perceived that AdaptNet, UNet++, UNet++\slash DS, and FEDNet have achieved the top four segmentation results. However, AdaptNet has achieved the highest mean IoU and Dice coefficient compared to the rival approaches. In particular, the proposed approach achieves 3.48\% improvement in mean IoU and 2.22\% improvement in mean Dice for lens segmentation compared to the best rival approach (UNet++). Moreover, the smaller standard deviation of IoU (10.56\% vs. 12.34\%) and Dice (8.56\% vs. 9.65\%) for AdaptNet compared to UNet++ confirms the reliability and effectiveness of the proposed architecture. For pupil segmentation, AdaptNet shows subtle improvement over the best rival approach (UNet++) regarding mean IoU and Dice while showing significant improvement regarding the standard deviation of IoU (1.91 vs. 4.05). Table~\ref{tab:ablation} provides an ablation study of AdaptNet. We have listed the Dice and IoU percentage with two different learning rates by gradually adding the proposed modules and blocks (for lens segmentation). It can be perceived from the results that regardless of the learning rate, each distinctive module and block has a positive impact on segmentation performance. We cannot test the FFD block separately since it is bound with the SSF module.

\begin{table}[t!]
\caption{Impact of different modules on the segmentation results of AdaptNet.}
\label{tab:ablation}
\centering
\begin{tabular}{P{1.2cm} P{1.2cm} P{1.2cm} P{1.2cm} P{1.2cm} P{1.2cm} P{1.2cm} P{1.2cm}}
\specialrule{.12em}{.05em}{.05em}
\multicolumn{4}{c}{Components}&\multicolumn{2}{c}{lr = 0.001}&\multicolumn{2}{c}{lr = 0.002}\\\cmidrule(lr){1-4}\cmidrule(lr){5-6}\cmidrule(lr){7-8}
Baseline&SSF&SHA&CPF&IoU(\%)&Dice(\%)&IoU(\%)&Dice(\%)\\\specialrule{.12em}{.05em}{.05em}
\checkmark&\xmark&\xmark&\xmark&82.79&89.94&84.33&90.90\\
\checkmark&\checkmark&\xmark&\xmark&83.54&90.33&84.99&91.22\\
\checkmark&\checkmark&\checkmark&\xmark&84.76&91.12&86.34&92.17\\
\rowcolor{lightgray} \checkmark&\checkmark&\checkmark&\checkmark&85.03&91.28&87.09&92.62\\
\specialrule{.12em}{.05em}{.05em}
\end{tabular}
\end{table}

\begin{figure}[!tb]
    \centering
    \includegraphics[width=0.97\textwidth]{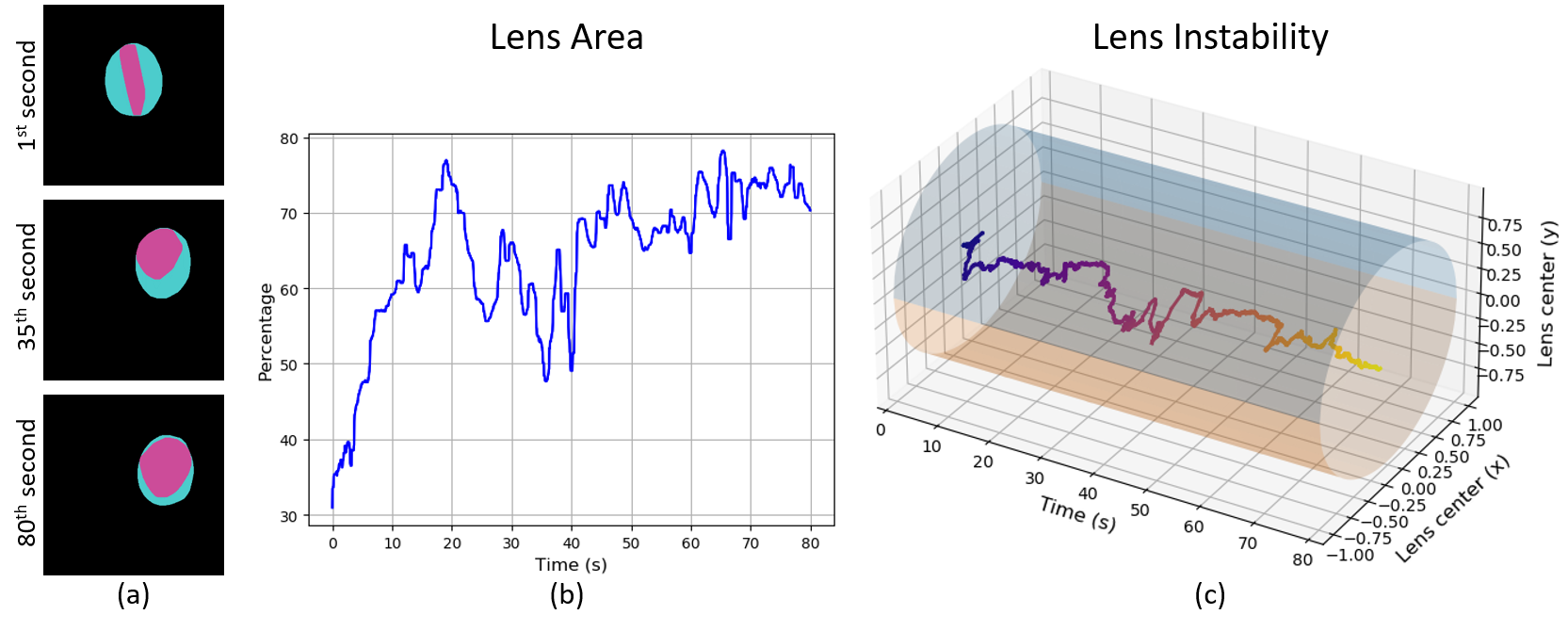}
    \caption{The lens statistics for one representative cataract surgery video.}
    \label{fig:ID}
\end{figure}

Fig.~\ref{fig:ID} shows the post-processed lens segments (pink) and pupil segments (cyan) from a representative video in different time slots (a), the relative lens area over time (b), and relative lens movements over time (c). Due to lens instability, a part of the lens is sometimes placed behind the iris, as shown in the segmentation results in the 35th second. Accordingly, the visible area of the lens can change independently of the unfolding state. Hence, the relative position of the lens should also be taken into account for lens unfolding delay computations. As can be perceived, the visible area of the lens is near maximum at 20 seconds after the implantation phase, and the lens is located nearly at the center of the pupil at this time. Therefore, the lens unfolding delay is 20 seconds in this case. However, the lens is quite unstable until 70 seconds after implantation. 
\section{Conclusion}
\label{sec: Conclusion}
Lens irregularity detection is a highly relevant problem in ophthalmology, which can play a prominent role in predicting and preventing lens relocation after surgery. This paper focuses on two significant steps towards lens irregularity detection: (i) ``lens implantation phase" detection and (ii) lens/pupil segmentation. In particular, We propose an end-to-end recurrent convolutional network that can extract spatio-temporal dependencies to detect the lens implantation phase accurately. Moreover, we propose a novel semantic segmentation network termed as AdaptNet. The proposed approach can deal with severe deformations and scale variations in the intraocular lens by adaptively fusing sequential and parallel feature maps. Experimental results reveal the effectiveness of the proposed phase recognition and semantic segmentation networks.

\bibliographystyle{splncs04}
\bibliography{bibtex.bib}

\newpage
\section{Supplementary Material}
\beginsupplement
Table~\ref{tab:specification} lists the specifications of the rival state-of-the-art approaches used in our evaluations. In ``Upsampling" column, ``Trans Conv" stands for \textit{Transposed Convolution}.

\begin{table*}[h!]
\renewcommand{\arraystretch}{0.9}
\caption{Specifications of the proposed and rival segmentation approaches.}
\label{tab:specification}
\centering
\begin{tabular}{lcccccccccc}
\specialrule{.12em}{.05em}{.05em}
Model && Backbone && Params && Upsampling && Reference && Year\\\specialrule{.12em}{.05em}{.05em}
UNet$++$ (\slash DS) &&VGG16&&24.24 M&& Bilinear &&~\cite{UNet++} && 2020\\
MultiResUNet &&\xmark&& 9.34 \enspace M&& Trans Conv &&~\cite{MultiResUNet} && 2020\\
CPFNet &&ResNet34&&34.66 M&& Bilinear &&~\cite{CPFNet} && 2020\\
dU-Net &&\xmark &&31.98 M&&Trans Conv &&~\cite{dU-Net} && 2020\\
CE-Net &&ResNet34&&29.90 M&&Trans Conv&&~\cite{CE-Net} && 2019\\
FED-Net &&ResNet50&&59.52 M&& Trans Conv \& PixelShuffle &&~\cite{FED-Net} && 2019\\
PSPNet &&ResNet50&&22.26 M&&Bilinear&&~\cite{PSPNet} && 2017\\
SegNet &&VGG16&&14.71 M&&Max Unpooling&&~\cite{SegNet} && 2017\\
U-Net &&\xmark &&17.26 M&& Bilinear &&~\cite{U-Net} && 2015\\\cdashline{1-11}[0.8pt/1pt]
AdaptNet&&VGG16 &&23.61 M&& Bilinear &&\multicolumn{3}{c}{Proposed}\\
\specialrule{.12em}{.05em}{0.05em}
\end{tabular}

\end{table*}

\begin{figure}[!h]
    \centering
    \includegraphics[width=1\columnwidth]{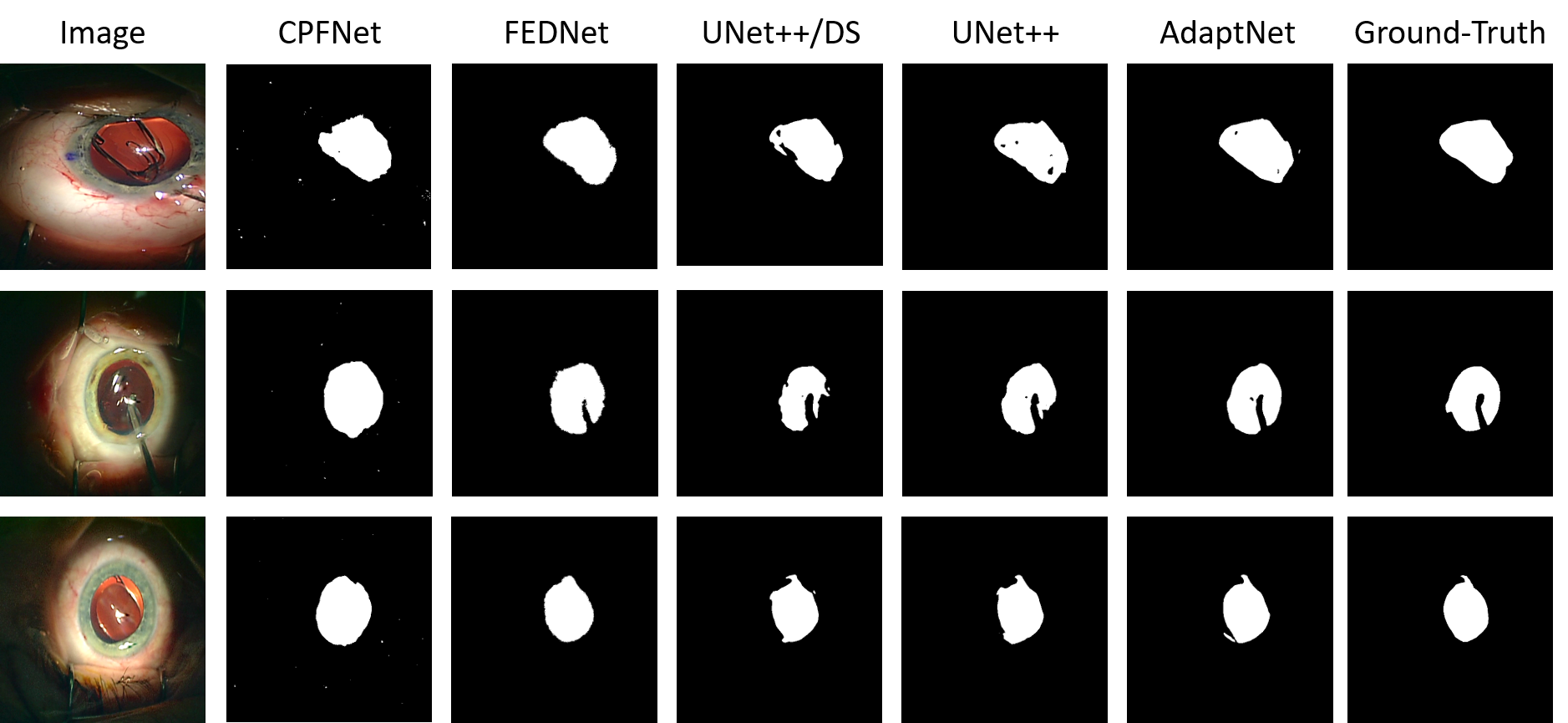}
    \caption{Qualitative comparisons among the top five segmentation approaches.}
    \label{fig: qualitative}
\end{figure}

Fig.~\ref{fig: qualitative} presents qualitative comparisons among the top five approaches for lens segmentation in three representative frames. It can be perceived from the figure that AdaptNet can provide the most visually close segmentation results to the ground truth. Moreover, AdaptNet is more robust against lens deformations as it provides the most delineated predictions compared to the rival approaches.

\begin{figure}[t]
    \centering
    \includegraphics[width=1\columnwidth]{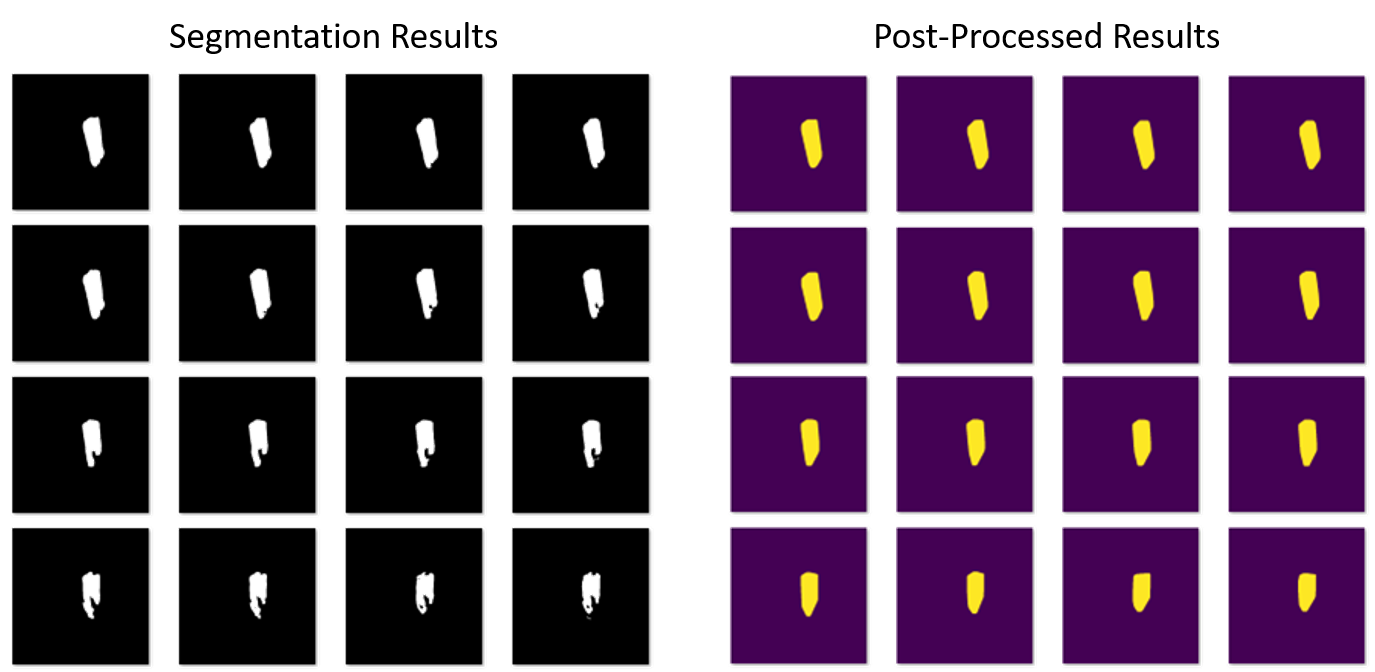}
    \caption{Qualitative comparisons among the top five segmentation approaches.}
    \label{fig: post-proc}
\end{figure}

Fig.~\ref{fig: post-proc} demonstrates the effect of post-processing on segmentation results. We use three morphological operations to improve the semantic segmentation results: (i) opening (with the kernel size of $10\times 10$) to attach the separated regions due to instrument covering, (ii) closing (with the kernel size of $15\times 15$) to remove the distant wrong detections, and (iii) convex polygon.  Since instruments usually cover a part of the pupil and intraocular lens during surgery, the segmentation results may contain some holes in the location of instruments. However, the occluded parts should be included in the lens and pupil area. Since the pupil is inherently a convex object, and the intraocular lens is usually convex during unfolding, we draw the smallest convex polygon around these objects to retrieve the occluded segments. For convex polygons, we used the ``Scipy ConvexHull" function.

\end{document}